\documentclass[journal]{IEEEtran}
\makeatletter
\newcommand\showlayout{%
  \typeout{Page width: \the\paperwidth}%
  \typeout{Page height: \the\paperheight}%
  \typeout{Text width: \the\textwidth}%
  \typeout{Text height: \the\textheight}%
  \typeout{Line w: \the\linewidth}%
  \typeout{Column w: \the\columnwidth}%
}
\makeatother

\usepackage{amsmath}
\usepackage{mathtools}
\usepackage{algpseudocode}
\usepackage{xcolor}
\usepackage{bm}
\usepackage{amssymb} 
\usepackage{siunitx}
\usepackage{enumitem}
\usepackage{mathtools}
\usepackage{placeins}
\usepackage{booktabs}
\usepackage{makecell}
\usepackage{microtype}
\usepackage{lipsum}
\usepackage{optidef}
\usepackage{eso-pic}
\usepackage{optidef}
\usepackage{svg}

\usepackage{balance}
\usepackage{amsmath,amsfonts}
\usepackage{array}
\usepackage{textcomp}
\usepackage{stfloats}
\usepackage{url}
\usepackage{verbatim}
\usepackage{graphicx}
\usepackage[export]{adjustbox}
\usepackage[linesnumbered,ruled,vlined]{algorithm2e}
\usepackage{balance}
\usepackage{xcolor}
\usepackage{setspace}
\usepackage{amsthm} 

\usepackage[switch,modulo]{lineno}

\usepackage[breaklinks]{hyperref} 
\usepackage[capitalise]{cleveref}
\usepackage{tikz}

\hypersetup{
    colorlinks,
    linkcolor={red!80!black},
    citecolor={blue!90!black},
    urlcolor={blue!90!black}
}

\def\BibTeX{{\rm B\kern-.05em{\sc i\kern-.025em b}\kern-.08em
    T\kern-.1667em\lower.7ex\hbox{E}\kern-.125emX}}

\long\def\hide#1{}

\NewDocumentCommand{\grad}{e{_^}}{%
  \mathop{}\!
  \mathop{}\mspace{-1mu}
  \nabla
  \IfValueT{#1}{_{\!#1}}
  \IfValueT{#2}{^{#2}}
  \mspace{-1mu}
}

\newcommand\scalemath[2]{\scalebox{#1}{\mbox{\ensuremath{\displaystyle #2}}}}

\let\ORGhypersetup\hypersetup
\protected\def\hypersetup{\ORGhypersetup}
\newtheorem{theorem}{Theorem}

\makeatletter
\def\th@plain{%
  \thm@notefont{}
  \itshape 
}
\def\th@definition{%
  \thm@notefont{}
  \normalfont 
}
\makeatother

\usepackage[font=small,position=bottom]{caption}
\usepackage[font=footnotesize,position=bottom]{subcaption}

\DeclareMathSymbol{\shortminus}{\mathbin}{AMSa}{"39}
\newcommand{\medminus}{\scalebox{0.6}[0.7]{\(-\)}}
\newcommand{\minus}{\mathchoice{-}{-}{\medminus}{\shortminus}}

\newcommand{\Del}{\mathop{}\!\Delta}

\DeclarePairedDelimiter\brackets{[}{]}
\DeclarePairedDelimiter\ceil{\lceil}{\rceil}

\DeclareMathOperator{\expectedValue}{\mathbb{E}}
\newcommand{\E}[1]{\expectedValue\brackets{#1}}

\renewcommand{\v}[1]{\bm{#1}}  

\newcommand{\etal}{\textit{et al}., }
\newcommand{\ie}{i.e., }
\newcommand{\eg}{\textit{e}.\textit{g}.\ }

\def\cidxth{p}
\def\csizeth{P}
\def\cidxest{q}
\def\csizeest{Q}  
\def\cSetTheory{\mathcal{A}_{\text{CS}}}  
\def\cSetEst{\hat{\mathcal{A}}}  

\def\cohThreshold{\gamma_{\text{min}}}

\def\noisePow{\sigma_v^2}
\def\targetPow{\sigma_s^2}
\def\interfPow{\sigma_m^2}
\def\interfCov{\v{S}_m}

\def\FourierPair{\overset{\scriptscriptstyle \mathcal{F}}{\longleftrightarrow}}
\def\xmod{{\bar x}_{[N]}^{(\alpha_{\cidxth})}} 

\def\ModSet{\mathcal{M}} 
\def\modf{\mu} 
\def\CandSet{\mathcal{C}} 

\def\cycSize{C_k}  
\def\cidx{c}  
\def\cmax{C_{\text{max}}}  
\def\ResNoise{\eta}

\def\noiseCorrTh{\nu}  
\def\noiseCorr{\kappa}  

\newcommand{\proc}[1]{\{{#1}\}}  
\newcommand{\fourier}[1]{#1}  
\newcommand{\fourierth}[1]{#1}  

\newcommand{\xfourier}[0]{\fourierth{x}}

\usetikzlibrary{positioning, arrows.meta, shapes.geometric}
\usetikzlibrary{calc}

\title{Cyclic MPDR Beamforming for Suppression of Almost-Cyclostationary Acoustic Interference}
\author{Giovanni Bologni, Martin Bo Møller, Richard Heusdens and Richard C.~Hendriks 
\thanks{This work was partly supported by the Dutch Research Council (NWO)
and partly by the Signal Processing Systems Group, Delft University of
Technology, Delft, The Netherlands. 
Giovanni Bologni, Richard Heusdens, and Richard C. Hendriks are with the Faculty of Electrical Engineering, Mathematics, and
Computer Science, Delft University of Technology, The Netherlands (e-mails: G.Bologni@tudelft.nl; R.Heusdens@tudelft.nl; R.C.Hendriks@tudelft.nl).
Richard Heusdens is also with Netherlands Defence Academy, 1781 AC Den Helder, The Netherlands.
Martin Bo Møller is with Bang \& Olufsen, Struer, 7600, Denmark (e-mail: acoustics.moeller@gmail.com),
Corresponding author: Giovanni Bologni.}}
\begin{document}
\showlayout
\pagestyle{plain}
\pagenumbering{arabic}
\maketitle

\begin{abstract}
Conventional acoustic beamformers typically assume short-time stationarity and process frequency bins independently, ignoring inter-frequency correlations.
This is suboptimal for almost-periodic noise sources such as engines, fans, and musical instruments: these signals are better modeled as (almost) cyclostationary (ACS) processes with statistically correlated spectral components.
This paper introduces the cyclic minimum power distortionless response (cMPDR) beamformer, which extends the conventional MPDR to jointly exploit spatial and spectral correlations.
Building on frequency-shifted (FRESH) filtering, it suppresses noise components that are coherent across harmonically related frequencies, reducing residual noise beyond what spatial filtering alone can achieve.
To address inharmonicity, where partials deviate from exact integer multiples of a fundamental frequency, we estimate resonant frequencies from a periodogram and derive frequency shifts from their pairwise spacing.
Theoretical analysis yields closed-form expressions for residual noise and proves that output power decreases monotonically with the number of cyclic components.
Experiments on synthetic harmonic noise and real UAV motor recordings confirm these findings: in low-SNR scenarios, the cMPDR achieves up to \SI{5}{\decibel} improvement in SI-SDR over the MPDR, yields consistent STOI gains, and remains effective with a single microphone.
When spectral correlation is absent, the method reduces to conventional MPDR and does not degrade performance.
These results suggest that cyclic processing is a viable direction for acoustic noise reduction that deserves further investigation.
Code and audio samples are available at \url{https://github.com/Screeen/cMPDR}.
\end{abstract}
\FloatBarrier

\section{Introduction}
\label{sec:intro}
\IEEEPARstart{R}{ecovering} a weak target signal in acoustic environments dominated by structured, periodic noise remains a challenge in acoustic signal processing and for industrial communication systems~\cite{zaki_method_2025}.
{
Conventional single- and multi-channel speech enhancement methods, including approaches based on deep neural networks (DNNs)~\cite{luo_fasnet_2019,tesch_insights_2023,quan_spatialnet_2024,shetu_ultra_2024} and classical power spectral density (PSD) techniques such as spectral subtraction or Wiener filtering, often struggle in extreme low signal-to-noise ratio (SNR) regimes where the target is entirely masked by high-energy interference~\cite{hao_unetgan_2019,schulz_effects_2026}.
Additionally, DNNs cannot easily be constrained to preserve a signal from a target direction~\cite{cohen_explainable_2025}, and they can be sensitive to training data.
}

Spatial filtering, exemplified by the minimum power distortionless response (MPDR) beamformer, offers a partial solution~\cite{capon_high-resolution_1969,gannot_consolidated_2017}.
The MPDR minimizes received power while preserving the signal from a target direction~\cite{doclo_acoustic_2010,gannot_signal_2001}.
Standard implementations in the short-time Fourier transform (STFT) domain assume the signal is wide-sense stationary (WSS), implying that different frequency bins are uncorrelated~\cite{napolitano_-_2020}.
\begin{figure}[!tb]
    \centering
    \includegraphics[width=\linewidth]{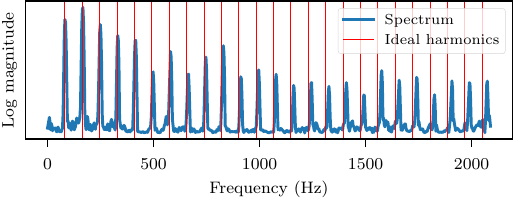}
    \caption{Spectrum of a cello note ($\text{E2}\approx\SI{82}{\hertz}$) with ideal harmonics. Higher harmonics deviate from integer multiples of the fundamental.
    }
    \label{fig:cello_harmonics_deviation}
\end{figure}

However, acoustic noise from rotating machinery, drones, and ventilation fans, 
exhibits periodic fluctuations better described by the cyclostationary (CS), or periodically correlated, model~\cite{gardner_statistical_1986,gardner_cyclostationarity_1994,feher_short_1995,gardner_cyclostationarity_2006,antoni_cyclostationarity_2009}.
The key distinction between CS and WSS processes lies in their frequency-domain behavior: 
CS signals possess statistical dependencies between harmonically related frequency bins.
Frequency-shift (FRESH) filters can exploit these spectral correlations to separate CS signals that overlap in both time and frequency, provided they possess distinct cyclic frequencies, which are defined in~\cref{ssec:signal_model}~\cite{gardner_cyclic_1993}.
These filters have long been used in telecommunications to localize and extract target CS signals from noise~\cite{schell_progress_1990,chevalier_constrained_1997,chevalier_blind_1998,zhang_blind_2001,zhang_reduced-rank_2006,du_class_2008}.
{More recently, the application of FRESH beamforming to mechanical fault localization has also been explored~\cite{zhang_localization_2023}.}
{
However, most FRESH-based approaches are developed for telecommunications and assume strict cyclostationarity, where cyclic frequencies are known exactly and reside on a uniform harmonic grid. 
Such rigid models are ill-suited for physical acoustic sources, which are more accurately characterized as almost-cyclostationary (ACS).}

In ACS processes, inharmonicity causes spectral correlations to deviate from the ideal harmonic series.
Physical mechanisms drive this inharmonicity:
in string instruments, finite stiffness causes higher partials to drift sharp~\cite{fletcher_normal_1964};
\cref{fig:cello_harmonics_deviation} illustrates this effect using a recording of a cello playing the note E2~\cite{iowa}.
In vibrating objects with multiple spatial degrees of freedom, such as drumheads, spectral components may occur at non-harmonic frequencies altogether.
Conventional cyclic methods that rely on exact harmonic structures fail when these deviations occur, as the expected correlation between bins vanishes \cite[Fig.~1d]{ojeda_adaptive-fresh_2010,bologni_cyclic_2025}.
This mismatch between CS model and ACS reality prevents traditional FRESH filters from effectively exploiting the spectral redundancy of the signal. 

In the acoustic domain, cyclostationarity has primarily been leveraged to model the target speech.
In the context of speech enhancement, Jensen \etal~\cite{jensen_harmonic_2020-1} focused on modeling speech as a harmonic signal with energy concentrated at integer multiples of a fundamental, while still assuming spectral uncorrelation.
Following this, a cyclic multichannel Wiener filter (cMWF) that exploits the CS property of voiced speech was proposed~\cite{bologni_cyclic_2025}.
However, these approaches face two limitations.
First, tracking the rapidly varying fundamental frequency and spectral statistics of speech is difficult.
Second, they rely on the strict harmonic model, where cyclic frequencies are exact integer multiples of a fundamental, which clashes with the slightly inharmonic reality of acoustic sources described above.
We therefore shift our attention from the highly varying target to the more stable, dominant interferer.
By exploiting the spectral redundancy of the ACS noise rather than the target, distortionless recovery becomes possible even when the target lacks periodic structure or is too variable to track reliably.
{\Cref{tab:cs_comparison} positions the cMPDR against prior cyclostationary methods. }

In this paper, we propose the cyclic minimum power distortionless response (cMPDR) beamformer, a model-based algorithm for cancelling dominant ACS interferers.
By augmenting the observation vector with frequency-shifted versions of the input, the cMPDR exploits the statistical redundancy of the interferer to steer nulls in both space and frequency, while providing distortionless guarantees for the target signal.
This allows for the recovery of weak signals even in single-channel configurations where spatial discrimination is impossible.
To handle the challenges of inharmonicity and partial correlation inherent to ACS signals, we introduce two technical innovations: a \textit{difference-based shift calculation} that adapts to the true spectral support of the noise instead of assuming a perfect harmonic series, and a \textit{coherence-based shift selection} that prunes weakly correlated frequency-shifted components, limiting computational complexity and letting the method degrade gracefully to the classical MPDR when spectral redundancy is weak or absent.

Theoretical analysis yields closed-form expressions for the residual noise power and output SNR, showing that the noise reduction capability increases with the spectral correlation of the interferer and proving that adding cyclic components does not degrade performance.
Experiments on synthetic harmonic noise, real UAV motor recordings (DREGON dataset), and non-stationary noise (Freesound/MUSAN dataset) confirm these predictions, with scale-invariant signal-to-distortion-ratio (SI-SDR) gains of up to \SI{5}{\decibel} over MPDR in low-SNR conditions (\eg $\text{SNR} \leq \SI{-10}{\decibel}$) and consistent intelligibility improvements.
Under non-stationary noise without persistent spectral correlation, cMPDR reduces to MPDR, exhibiting identical performance.

Our contributions are as follows.
\begin{itemize}
    \item We introduce the cMPDR, a beamformer that exploits spectral and spatial correlations of the dominant noise rather than the target, enabling distortionless recovery even when the target is buried in noise.
    \item We derive closed-form expressions for the achievable subband SNR gain as a function of the spectral correlation of the interferer, quantifying the advantage of cMPDR over spatial-only beamforming.
    \item We propose a difference-based shift calculation that derives candidate frequency shifts from the spacing between the estimated resonant frequencies, together with a coherence-based shift selection that retains only strongly correlated frequency-shifted components.
    \item We publicly release the Python implementation of the proposed method~\cite{github_2026}, along with online audio demos.
\end{itemize}

\let\mc\makecell
\begin{table*}[!t]
\caption{Comparison of representative cyclostationary-based methods. Most prior work was developed for telecommunications or mechanical signal analysis, whereas only limited work exists in audio signal processing. $^*$: this work is about localization.}
\label{tab:cs_comparison}
\small
\centering
\begin{tabular}{lcccccc}
\toprule
Method & \mc{Application\\ domain} & \mc{Cyclostationarity\\ exploited in} & \mc{Supports unknown\\ cyclic frequencies} & \mc{Rejects weakly correlated\\cyclic components} & \mc{Multi-\\ channel}
& \mc{Distortion-\\ less} \\
\midrule
FRESH~\cite{gardner_cyclic_1993} & Comm.& Target+noise & No & No & No & No \\
Cyclic MUSIC$^*$~\cite{schell_progress_1990} & Comm.&Target & Yes & No & No & No \\
Cyclo BF$^*$~\cite{zhang_localization_2023} & Mech. & Target & No & No & Yes & No \\
BA-FRESH~\cite{zhang_blind_1999-1}& Comm.& Target & No & No & No & No \\
FRESH BF~\cite{zhang_reduced-rank_2006,du_class_2008}& Comm.& Target & No & No & Yes & No \\
FRESH-LCMV~\cite{chevalier_constrained_1997,chevalier_blind_1998}& Comm.& Target & No & No & Yes & Yes \\
Adaptive-FRESH~\cite{ojeda_adaptive-fresh_2010}& Comm. & Target & No & No & No & No \\
cMWF~\cite{bologni_cyclic_2025}& Audio & Target & No & No & Yes & No \\
\textbf{cMPDR (proposed)} & \textbf{Audio} & \textbf{Noise} & \textbf{Yes} & \textbf{Yes} & \textbf{Yes} & \textbf{Yes} \\
\bottomrule
\end{tabular}
\end{table*}

\section{Background}\label{sec:background}\noindent
\subsection{Cyclostationary processes}\label{ssec:signal_model}\noindent
This section introduces the definition of CS processes and their spectral representation.
Time-domain random processes are denoted by a bar.
A real-valued discrete-time random process $\proc{{\bar x}(n), n \in \mathbb{Z}}$ is wide-sense \textit{cyclostationary} if both its mean and covariance function are periodic with some integer period $\csizeth$:
\begin{align}
    m_{\bar x}(n) &= m_{\bar x}(n + \csizeth),\\
    r_{\bar x}(n, \tau) &= r_{\bar x}(n + \csizeth, \tau),
    \quad \forall n, \tau \in \mathbb{Z}.
\end{align}
{In other words, the statistical properties of the process repeat every $\csizeth$ samples.}
As the mean and the covariance of a CS process are periodic in $n$ with period $\csizeth$, they admit a Fourier series expansion over a set of \emph{cyclic frequencies}
\begin{align}
    \scalemath{1.}{
    \cSetTheory = \{\alpha_{\cidxth} = 2 \pi \cidxth / \csizeth\}_{\cidxth=0}^{\csizeth-1}
    }.
\end{align}
{These are the frequencies at which the statistics oscillate.}
A cyclic frequency $\alpha_{\cidxth}$ is also referred to as a \emph{resonant frequency} of the process.
By expanding $r_{\bar x}(n,\tau)$ in a Fourier series with respect to $n$ and applying a discrete-time Fourier transform with respect to $\tau$, we get a function $S_x(\alpha_{\cidxth}, \omega)$ of two frequency variables: the
\emph{cyclic} frequency $\alpha_{\cidxth}$ and the \emph{spectral} frequency $\omega$ (assuming convergence of the infinite sum):
\begin{align}
    S_x(\alpha_{\cidxth}, \omega) = \sum_{\tau = -\infty}^{\infty} \sum_{n=0}^{\csizeth-1} r_{\bar x}(n, \tau) e^{-j (\alpha_{\cidxth} n + \omega \tau)}.
\end{align}
This quantity is known as spectral correlation density (SCD), or cyclic spectrum, as for finite-length processes
\footnote{
A process with support other than $(-\infty, \infty)$ cannot be CS.
Yet, if the process exhibits CS statistics over an interval much longer than the one relevant to the application, it can be reasonably treated as CS within that interval, in analogy to the standard practice of treating finite-duration signals as WSS~\cite{gallager_principles_2007}.
}
it is also given by~\cite{gardner_cyclostationarity_1994}:
\begin{align}\label{eq:spec_corr_freq}
    S_x(\alpha_{\cidxth}, \omega_k) = \E{\xfourier(\omega_k) \xfourier^*(\omega_k - \alpha_{\cidxth})},
\end{align}
where $\xfourier(\omega_k)$ denotes the process given by the $K$-point discrete Fourier transform (DFT) of $\proc{{\bar x}_{[K]}(n), n=0,\ldots,K-1}$,
\begin{align}
\xfourier(\omega_k) = \sum_{n=0}^{K-1} \bar{x}_{[K]}(n)e^{-j\omega_k n}.
\end{align}
{The SCD in~\cref{eq:spec_corr_freq} measures the correlation between the frequency component at $\omega_k$ and the component at $\omega_k - \alpha_{\cidxth}$, that is separated by the cyclic frequency $\alpha_{\cidxth}$.}
The SCD boils down to the conventional power spectral density (PSD) $S_x(\omega_k)$ for $\alpha_\cidxth = 0$:
\begin{align}\label{eq:cyc_spec_0_theory}
    S_x(0, \omega_k) = S_x(\omega_k) = \E{|\xfourier(\omega_k)|^2}. 
\end{align}
In other words, when $\alpha_\cidxth = 0$, we recover the standard PSD, which measures power at each frequency without considering correlations between different frequencies.

We also introduce a normalized version of the cyclic spectrum, called spectral coherence, which is given by~\cite{gardner_exploitation_1991, antoni_detection_2012}:
\begin{align}\label{eq:coherence_single_ch}
    \gamma_{x}(\alpha_{\cidxth}, \omega_k) = \frac
    {|S_{x}(\alpha_{\cidxth}, \omega_k)|^2}
    {S_{x}(\omega_k) S_{x}(\omega_k -\alpha_{\cidxth})},
\end{align}
where $0 \leq \gamma_{x}(\alpha_{\cidxth}, \omega_k) \leq 1$.
The normalization makes the spectral coherence easier to interpret: values close to 1 indicate strong correlation between the frequency components.

A key property of CS processes is to exhibit inter-frequency correlations that can be measured by the cyclic spectrum or the spectral coherence.
In fact, $\xfourier(\omega_1)$ is correlated with $\xfourier(\omega_2)$ for $|\omega_1 - \omega_2| = \alpha_{\cidxth}$, for every $\alpha_{\cidxth} \in \cSetTheory$.
Intuitively, if we measure $\xfourier(\omega_1)$, we know something about $\xfourier(\omega_2)$.
In contrast, the spectral components of WSS processes are asymptotically uncorrelated. For example, for white Gaussian noise, we have that
$S_x(\alpha_{\cidxth}, \omega_k) = \gamma_{x}(\alpha_{\cidxth}, \omega_k) = 0$ for all $\alpha_{\cidxth} \neq 0$.
{This means that for WSS processes, knowing the power at one frequency does not tell us anything about the power at another frequency.}
Notice that all quantities in this section are defined for a single random process, but generalizing the notions to the cross-statistics between multiple processes is straightforward.

Real-world acoustic signals are often not strictly periodic and are therefore more appropriately modeled as ACS processes~\cite{madisetti_cyclostationary_2009}.
In contrast to strictly CS models, the cyclic frequencies are not restricted to integer multiples of a single fundamental but may take arbitrary values, leading to non-uniform spectral spacing.
We define the cyclic set
\begin{align}
\mathcal{A}_{\text{ACS}} = \mathcal{A} = \{\alpha_{\cidxth}\}_{\cidxth=0}^{\csizeth-1}
\end{align}
which contains all cyclic frequencies at which second-order spectral correlations occur.
In the acoustic scenarios discussed in the introduction, $\mathcal{A}$ is generally inharmonic and unknown, and will be estimated from data in \Cref{ssec:freq_est_period}.
For a detailed treatment of the theory of ACS processes, see \cite[Sec.~3.2]{gardner_cyclostationarity_2006}.

\subsection{Estimation of the cyclic spectrum}
\label{ssec:est_spec_corr_acp}\noindent
The cyclic spectrum, defined in~\cref{eq:spec_corr_freq}, is given by the ensemble expectation of the product of the signal at frequencies $\omega_k$ and $\omega_k - \alpha_{\cidxth}$.  
In the beamforming context considered here, the SCD is estimated by a finite temporal average over data.
Next, since $\omega_k$ is by definition a DFT sampling point but $\omega_k - \alpha_{\cidxth}$ generally is not, the frequency shift must be handled explicitly.  
One could approximate it by increasing the DFT size to obtain a finer resolution, but this would be computationally inefficient.  
A more practical approach is to modulate the signal in the time domain by $e^{j \alpha_{\cidxth} n}$, which shifts the spectrum prior to applying a standard DFT.  
The resulting \emph{time-averaged cyclic periodogram} (ACP) provides a widely used estimator of the cyclic spectrum~\cite{gardner_measurement_1986}. 
The ACP coincides with the Welch PSD estimator for $\alpha_\cidxth = 0$~\cite{antoni_cyclic_2007}, and under mild regularity conditions it yields consistent SCD estimates even from a single record~\cite{gardner_measurement_1986}.
Other methods for SCD estimation may offer faster computations if knowledge of the cyclic spectrum at all spectral and cyclic frequencies is required~\cite{roberts_computationally_1991,borghesani_faster_2018,alsalaet_fast_2022}.
However, since we require the SCD only for a limited set of cyclic frequencies, the ACP is well suited to our needs.

To implement the ACP, we first modulate the signals in the time-domain, then process them in the STFT domain.
Let $\proc{{\bar x}(n), n \in \mathbb{Z}}$ and $\proc{{\bar y}(n), n \in \mathbb{Z}}$ be random processes sampled with sampling frequency $f_s$, and define $\proc{{\bar x}_{[N]}(n)}$ and $\proc{{\bar y}_{[N]}(n)}$ to equal $\proc{{\bar x}(n)}$ and $\proc{{\bar y}(n)}$ for $n=0, \ldots, N-1$ and zero elsewhere.
Using a DFT with size $K$ and hop size $R$, we obtain $L = \ceil{1 + (N - K) / R}$ STFT frames, where $\ceil{\cdot}$ is the ceiling function.
The DFT length determines the spectral resolution $\Del \omega \approx f_s / K~[\si{\hertz}]$.
By contrast, the spacing between resolvable cyclic frequencies follows from the total number $LR$ of observed samples, $\Del \alpha \approx f_s / (L R)~[\si{\hertz}]$, even though we only evaluate the SCD for a small number of cyclic shifts~\cite{gardner_measurement_1986}.
As mentioned earlier, these cyclic shifts do not generally align with the DFT bin spacing of $1/K$.
Rather than extracting $\fourierth{x}(\omega_k - \alpha_{\cidxth})$ by shifting the spectrum, we achieve the shift by modulating the time-domain signal with $e^{j \alpha_{\cidxth} n}$ before taking the DFT, exploiting the modulation property:
\begin{align}
\scalemath{1.}{
\fourierth{x}(\omega_k - \alpha_{\cidxth}) \FourierPair {\bar x}_{[N]}(n) e^{j \alpha_{\cidxth} n}.
}
\end{align}
The modulated signal in the time domain and its STFT counterpart are given by:
\begin{gather}
    \xmod(n) = {\bar x}_{[N]}(n) e^{j n \alpha_{\cidxth}}, \label{eq:modulation}\\
    \fourierth{x}(\omega_k - \alpha_{\cidxth}, \ell) = \sum_{n=0}^{K-1} \xmod(n + \ell R)w(n)e^{-j n \omega_k}, \label{eq:stft}
\end{gather}
where $\ell$ is the time-frame index and $w(n)$ is a window function with support in $\{0, \ldots, K-1\}$.
The ACP estimate at cyclic frequency $\alpha_{\cidxth}$ and spectral frequency $\omega_k$ is then given by:
\begin{align}\label{eq:acp_estimator}
    \hat{S}_{yx}(\alpha_{\cidxth}, \omega_k) = \frac{1}{L}\sum_{\ell=0}^{L-1} \fourierth{y}(\omega_k, \ell) \fourierth{x}^*(\omega_k - \alpha_{\cidxth}, \ell).
\end{align}
Likewise, the spectral coherence can be estimated as:
\begin{align}\label{eq:acp_estimator:coh}
    \hat{\gamma}_{yx}(\alpha_{\cidxth}, \omega_k) = \frac
    {|\hat{S}_{yx}(\alpha_{\cidxth}, \omega_k)|^2}
    {\hat{\tilde{S}}_{y}(\omega_k) \hat{\tilde{S}}_{x}(\omega_k -\alpha_{\cidxth})},
\end{align}
where the regularized PSDs are defined as \begin{equation*}
\hat{\tilde{S}}_{y}(\omega_k) = \max\{\hat{S}_{y}(\omega_k), \hat{S}_{y}^{\text{max}} / D_{\text{PSD}} \},~\hat{S}_{y}^{\text{max}} = \max_{\omega_k}{\hat{S}_{y}(\omega_k)},
\end{equation*} 
and similarly for $\hat{\tilde{S}}_{x}(\omega_k-\alpha_{\cidxth})$.
By constraining the PSDs to be at least $1/D_{\text{PSD}}$ of their peak value, this approach avoids divisions by near-zero values.

{
\subsection{Relation to cross-band filter and frequency leakage models}\label{ssec:leakage}\noindent
The spectral correlations defined in~\cref{ssec:signal_model} arise from the source itself: the periodicity of an ACS process induces statistical dependencies between harmonically related bins, as captured by the SCD in~\cref{eq:spec_corr_freq}.
Two other mechanisms produce cross-band correlations that are unrelated to source periodicity and should not be confused with the above.
First, convolution with a long room impulse response introduces cross-band coupling, modeled by cross-band filters and exploited through the convolutive transfer function (CTF) model~\cite{gilloire_adaptive_1992,avargel_system_2007,li_multichannel_2019-2}; this is a property of the acoustic system.
Second, STFT windowing causes frequency leakage, inducing spurious correlations between adjacent bins~\cite{chen_single-channel_2012, huang_minimum_2014}; this is an artifact of the processing chain.
Both effects are largely confined to bins in the immediate vicinity of the working frequency, whereas ACS links distant harmonic components across the full spectrum; the three families of methods are thus complementary rather than contradictory.
}

\subsection {Single-band beamforming}
\label{ssec:nb_beamform}\noindent
Following the overview of cyclostationary theory and methods, we now examine classical beamforming theory.
We denote matrices by bold capitals and vectors by bold lowercase letters.
Let $\v{x}(\omega_k, \ell) = [\fourier{x}_0(\omega_k, \ell)~\ldots~\fourier{x}_{M \minus 1}(\omega_k, \ell)]^T \in \mathbb{C}^M$ represent noisy and reverberant measurements from a $M$-elements microphone array in the STFT domain at time-frame $\ell$.
Hereafter, the time-frame index $\ell$ is omitted for simplicity.
The measurements are modeled as:
\begin{align}\label{eq:sig_mod_nb}
\v{x}(\omega_k) &=
\fourier{s}(\omega_k)\,{\v{a}}(\omega_k) + \v{v}(\omega_k)
= {\v{d}}(\omega_k) + \v{v}(\omega_k),
\end{align}
where
$
{\v{a}}(\omega_k) = \begin{bmatrix*}
    1&
    a_1(\omega_k)~
    \ldots~
    a_{M \minus 1}(\omega_k)
\end{bmatrix*}^T
$
is the relative transfer function (RTF) between a reference sensor and the remaining sensors,
$\fourier{s}(\omega_k)$ denotes the target signal at the reference microphone,
and $\v{v}(\omega_k)$ represents additive noise.
Without loss of generality, the first sensor is chosen as the reference.
The time-domain convolution of the signal and room impulse response (RIR) is approximated by a multiplication in the STFT domain, assuming the reverberation time is much shorter than a single STFT frame~\cite{talmon_relative_2009}.
The general goal of beamforming is to estimate a target signal as a linear combination of the noisy inputs.
A popular approach is the MPDR beamformer, which is formulated as the solution to:
\begin{mini}|s|[0]
    {\v{w}(\omega_k)}{\E{|\v{w}^H(\omega_k) \v{x}(\omega_k)|^2}}
    {}
    {\label{eq:mvdr_nb}}{}
    \addConstraint{\v{w}^H(\omega_k) \v{a}(\omega_k)}{= 1},
\end{mini}
where the goal is to reduce the power of the noisy signal as much as possible while retaining signals whose transfer function is $\v{a}(\omega_k)$.

\section{Proposed cMPDR beamformer}\label{sec:prop_algo}\noindent
This section presents the proposed cMPDR beamformer. 
The design extends the single-band model in~\cref{eq:sig_mod_nb} to a multi-band model, in which the received signal is evaluated at multiple, arbitrary frequencies, and the corresponding signal components are highly correlated.
In the following, we refer to these as ``frequency-shifted'' components, since evaluation at arbitrary frequencies is carried out by modulating the time-domain signal prior to the STFT (see~\cref{ssec:est_spec_corr_acp}).  

\begin{figure}[tb]
    \centering
    \subfloat[]{%
    \includegraphics[width=0.47\linewidth]{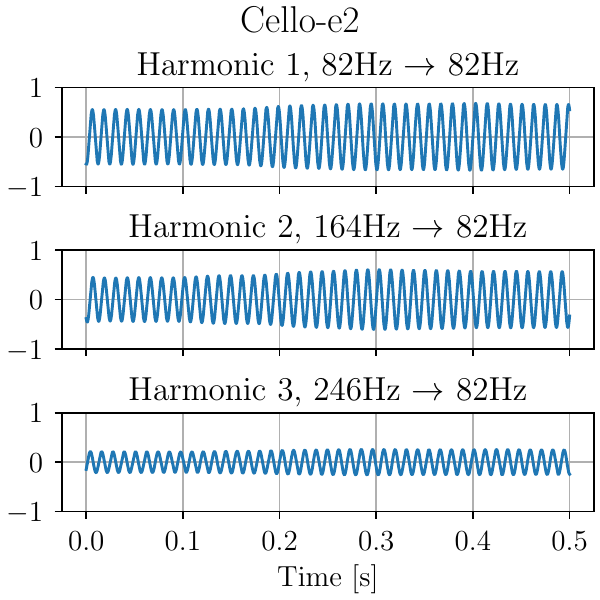}
    \label{fig:harmonic_components:cello}
    }
    \hfill
    \subfloat[]{%
    \includegraphics[width=0.47\linewidth]{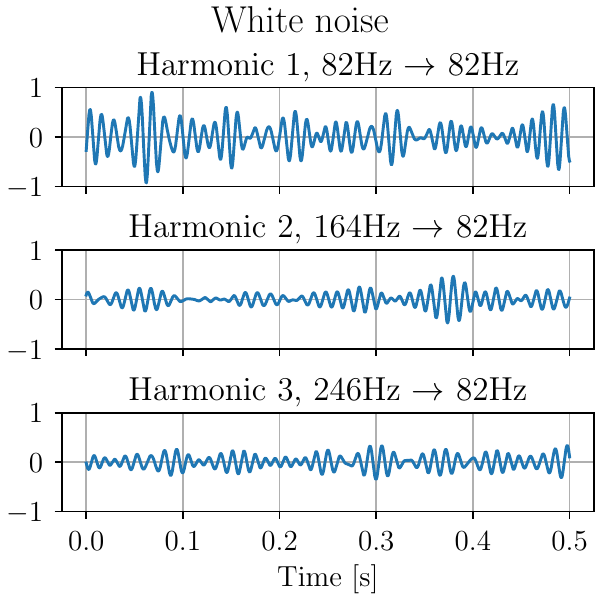}
    \label{fig:harmonic_components:wgn}
    }
    \caption{
    {
    Waveforms of cello (a) and white noise (b) after bandpass filtering around the first three harmonics and downshifting each component to \SI{82}{\hertz}, the fundamental frequency of the note.
    Each row corresponds to one frequency-shifted harmonic ($n=1,2,3$), forming the inputs to the cMPDR beamformer.
    The cello components are highly correlated, while the white noise components appear uncorrelated.
    }
    }
    \label{fig:harmonic_components}
\end{figure}

Before introducing the multi-band model, consider the example in~\cref{fig:harmonic_components} to build intuition on FRESH beamforming.
The same cello recording as in~\cref{fig:cello_harmonics_deviation} ($f_1\approx\SI{82}{\hertz}$) is used, as musical instrument recordings exhibit high spectral coherence.
The signal is decomposed into its first three harmonics by bandpass filtering around $n\cdot f_1$ for $n=1,2,3$. 
Each harmonic is then frequency-shifted (down-modulated) to $f_1$ and passed through a second bandpass filter centered at $f_1$.
The three rows in~\cref{fig:harmonic_components} correspond to these processed harmonics, \ie the signals obtained after downshifting each harmonic to the same frequency.
This sequence resembles, but does not exactly reproduce, the frequency-shift and STFT steps of the beamforming pipeline (see~\cref{fig:cmpdr_scheme}); the resulting signals are shown here in the time domain for visualization purposes only.
For the cello signal (\cref{fig:harmonic_components:cello}), the three downshifted components appear nearly identical up to a scaling and delay, indicating that they can be linearly combined.
This property is known as \emph{spectral redundancy}~\cite{adlard_frequency_2000}.
By contrast, applying the same steps to white noise (\cref{fig:harmonic_components:wgn}) yields components that exhibit no visible similarity and behave as independent realizations.

\subsection{Modulation sets \texorpdfstring{$\ModSet_k$}{}}\noindent
For each frequency bin $k$, the multi-band signal is formed by concatenating the received signal with its frequency-shifted versions.
The applied shifts are collected in the \textit{modulation set}
\begin{align}\label{eq:mod_set_definition}
\ModSet_k = \{\modf_{\cidx_k}\}_{\cidx_k = 0}^{\cycSize - 1},\qquad k = 0,\ldots,K-1,
\end{align}
where $\cycSize$ is the cardinality of the set, and $\modf_{\cidx_k}$ is a linear combination of the cyclic frequencies.
By convention, the first element is $\modf_{0_k} = 0$, corresponding to no modulation.
Depending on how the elements of $\ModSet_k$ are chosen, the multi-band model accommodates spectral correlations in either the target or the noise signal. 
In this work, the shifts are selected based on the resonant frequencies of the noise, so the noise is modeled as ACS while the target is assumed WSS, or possibly ACS with distinct cyclic frequencies.
In what follows, $\ModSet_k$ is assumed known; estimation is addressed in~\cref{sec:calc_mod_set}.

\subsection{Multi-band signal model}\noindent
In the multi-band model, the received signal and its shifted counterparts are stacked into the vector $\v{x}(\ModSet_k, \omega_k) \in \mathbb{C}^{M\cycSize}$, where $M$ is the number of microphones: 
\begin{multline}\label{eq:augmented_vector}
\v{x}(\ModSet_k, \omega_k) = \\
\begin{bmatrix*}
\v{x}(\omega_k) &
\v{x}(\omega_k \minus \modf_{1_k}) &
\cdots &
\v{x}(\omega_k \minus \modf_{\cycSize \minus 1})
\end{bmatrix*}^T,
\end{multline}
{For $M=1$, each element of this vector corresponds to one of the rows in~\cref{fig:harmonic_components:cello} (shown in the time domain for visualization): the first element is the signal at $\omega_k$ (row 1), the second is the signal shifted by $\modf_{1_k}$ (row 2), and so on.}
The modulated noise vector $\v{v}$ and reverberant signal vector $\v{d}$ are constructed similarly:
\begin{multline}\label{eq:augmented_vector_noise}
\v{v}(\ModSet_k, \omega_k) = \\
\begin{bmatrix*}
\v{v}(\omega_k) &
\v{v}(\omega_k \minus \modf_{1_k}) &
\cdots &
\v{v}(\omega_k \minus \modf_{\cycSize \minus 1})
\end{bmatrix*}^T.
\end{multline}
To simplify notation, we omit explicit dependencies on $\ModSet_k$ and $\omega_k$ from this point onward.
The modulated reverberant signal $\v{d}$ can be expressed as $\v{d} = \v{A}\v{s}$, where
$
\v{s} = \begin{bmatrix*} 
\fourier{s}(\omega_k)~ 
\ldots~
\fourier{s}(\omega_k - \modf_{\cycSize \minus 1}) 
\end{bmatrix*}^T \in \mathbb{C}^{\cycSize}
$
is the modulated signal at the reference microphone, and $\v{A} \in \mathbb{C}^{M\cycSize \times \cycSize}$ contains zero-padded, frequency-shifted RTFs. 
As an example, for the special case $\cycSize=2$, $\v{d}$ is given by:
\begin{align}\label{eq:reverberant_target}
\scalemath{0.98}{
\v{d} = \v{A}\v{s} = 
\begin{bmatrix*}
\v{a}(\omega_k) & \v{0}_{M(\cycSize-1)} \\
\v{0}_{M(\cycSize-1)} & \v{a}(\omega_k - \modf_{1_k})
\end{bmatrix*}
\begin{bmatrix*}
    \fourier{s}(\omega_k) \\
    \fourier{s}(\omega_k - \modf_{1_k})
\end{bmatrix*},
}
\end{align}
where $\v{0}_N$ represents a zero vector of size $N$.
The resulting multi-band signal model can now be expressed as:
\begin{align}\label{eq:sig_mod_wb}
\v{x} = \v{d} + \v{v} \in \mathbb{C}^{M\cycSize}.
\end{align}
Let us also define
\begin{align}\label{eq:spectral_cov_definition}
    \v{S}_{\v{x}}(\ModSet_k, \omega_k) = \v{S}_{\v{x}} = \E{\v{x}\v{x}^H} \in \mathbb{C}^{M\cycSize\times M\cycSize}
\end{align}
as the spectral-spatial covariance matrix across microphones and cyclic frequencies, where each entry corresponds to a cyclic spectrum.
In practice, the elements of $\v{S}_{\v{x}}$ are estimated using the ACP method detailed in~\cref{ssec:est_spec_corr_acp} per spectral frequency $\omega_k$, cyclic frequency $\modf_{\cidx_k} \in \ModSet_k$ and microphone pair.
Notice that, for $\ModSet_k=\{0\}$, we have that
\begin{align}\label{eq:cyc_spec_0}
    \v{S}_{\v{x}}(\{0\}, \omega_k) = \v{S}_{\v{x}}(\omega_k) \in \mathbb{C}^{M\times M},
\end{align}
\ie the spectral-spatial covariance matrix reduces to the spatial covariance matrix if only the null modulation is considered.
We additionally assume that the complex signals under analysis are \emph{proper}, meaning that their conjugate covariance vanishes, $\E{\v{x}\v{x}^T}=\v{0}$, and can thus be ignored hereafter~\cite{picinbono_circularity_1994,benesty_study_2010}.

\subsection{Beamformer design}\noindent
Based on the multi-band signal model, we design a beamformer that jointly processes microphones and frequency-shifted components.
Exploiting the spectral correlations captured in $\v{S}_{\v{x}}$, the beamformer minimizes the total output power across cyclic frequencies while enforcing a distortionless constraint for the desired RTF, analogous to the single-band MPDR.
The resulting cMPDR is obtained as the solution of the following optimization problem:
\begin{mini}|s|[0]
    {\v{w}}{\E{|\v{w}^H {\v{x}}|^2}}
    {}
    {}{}
    \addConstraint{\v{w}^H \v{a}_{0}}{= 1},
    \label{eq:cmpdr_problem}
\end{mini}
where $\v{a}_{0} =
\big[
\v{a}^T(\omega_k)~\v{0}_{M(\cycSize-1)}^T
\big]^T
= \v{A}\v{e}_1
$ 
is the target RTF padded with zeroes, corresponding to the first column of $\v{A}$, and $\v{e}_1$ is the standard basis vector of size $\cycSize$, with 1 at the first position and 0 elsewhere.
The solution to this problem is given by
\begin{align}\label{eq:cmpdr_solution}
    \v{w}_{\text{cMPDR}} = \frac{\v{S}_{\v{x}}^{-1}\v{a}_{0}}{\v{a}_{0}^H \v{S}_{\v{x}}^{-1} \v{a}_{0}}.
\end{align}
{To improve the conditioning of the estimated covariance matrix we employ diagonal loading (see~\cref{sec:exp_setup} for details).}
The multi-band noisy input $\v{x}$ contains the desired component $\v{d}(\omega_k)$, but also its modulated counterparts $\v{d}(\omega_k-\modf_{\cidx_k}),~\cidx_k=1,\ldots,\cycSize-1$, and the additive noise terms.

The constraint $\v{w}^H \v{a}_{0} = 1$ preserves only the non-modulated target at the reference microphone, $d_0(\omega_k)$.
The modulated target components are not explicitly constrained.
Introducing additional constraints, as in a linearly constrained minimum variance (LCMV) formulation, would require knowledge of the corresponding RTFs and reduce the achievable noise suppression~\cite{chevalier_constrained_1997}.
Instead, by minimizing the total output power $\E{|\v{w}^H\v{x}|^2}$, the proposed design treats the modulated target components as interference and suppresses them implicitly.
In the conventional single-band formulation, minimizing total output power (MPDR) and minimizing noise power, i.e., the minimum variance distortionless response (MVDR) criterion, lead to the same solution.
In the present multi-band formulation, this equivalence no longer holds: a cyclic MVDR, obtained by minimizing noise power only, would not penalize the modulated target components and would leave them at the output, whereas the proposed cMPDR suppresses them by minimizing total output power.

\section{Statistical analysis of the cMPDR}\label{sec:stat_props}\noindent
This section analyzes three statistical properties of the cMPDR beamformer.
For clarity, we focus on a single frequency bin $\omega_k$ and set $\cycSize=2$ resulting in $\ModSet_k = \{0, \modf_{1_k}\}$.
{
This simplified setting provides insight into the behavior of the cyclic beamformer.
First, we show that when the received signal is spectrally uncorrelated, the cMPDR reduces exactly to the classical MPDR.
Next, we compute the residual noise power and corresponding SNR improvements in closed form for $M=1$.
Finally, 
we show that in the general case the residual noise power decreases monotonically with $\cycSize$ when spectral correlation is present.
}

\subsection{Spectrally uncorrelated components}\label{ssec:stat:uncorr_components}\noindent
Consider a spectral component that is uncorrelated with its frequency-shifted counterpart.
If $\v{x}(\omega_k)$ is spectrally uncorrelated with $\v{x}(\omega_k - \modf_{1_k})$, then $\E{\v{x}(\omega_k)\v{x}(\omega_k - \modf_{1_k})^H} = \v{0}$, and the covariance matrix $\v{S}_{\v{x}}$ becomes block-diagonal.
The numerator of~\cref{eq:cmpdr_solution} simplifies to
\begin{align}\label{eq:example_cmpdr_uncorr}
\v{S}_{\v{x}}^{-1}\, \v{a}_0 =
\begin{bmatrix*}
\v{S}_{\v{x}}^{-1}(0, \omega_k)&\v{0}\\
\v{0}&\v{S}_{\v{x}}^{-1}(\modf_{1_k}, \omega_k)
\end{bmatrix*}
\begin{bmatrix*}
\v{a}(\omega_k) \\
\v{0}
\end{bmatrix*}
,
\end{align}
where we used the fact that the inverse of a block-diagonal matrix is a block-diagonal matrix whose blocks are the inverses of the original blocks.
Substituting~\cref{eq:cyc_spec_0} into~\cref{eq:example_cmpdr_uncorr}, and~\cref{eq:example_cmpdr_uncorr} into~\cref{eq:cmpdr_solution}, it follows that the cMPDR reduces to the MPDR: $\v{w}_{\text{cMPDR}} = [\v{w}_{\text{MPDR}}^T~\v{0}^T]^T$.
This implies that cyclic beamforming provides benefit only at frequencies where the noise exhibits spectral correlation, which typically occurs at the harmonic components.
{
Including additional frequency-shifted components $\v{x}(\omega_k - \modf_{2_k}), \v{x}(\omega_k - \modf_{3_k}), \ldots$ does not change the conclusion, provided that all such components are mutually uncorrelated and uncorrelated with $\v{x}(\omega_k)$.}

\subsection{Noise reduction performance: effects of correlation}\label{ssec:stat:noise_red_corr}
\begin{figure}[tb]
    \includegraphics[width=0.9\linewidth]{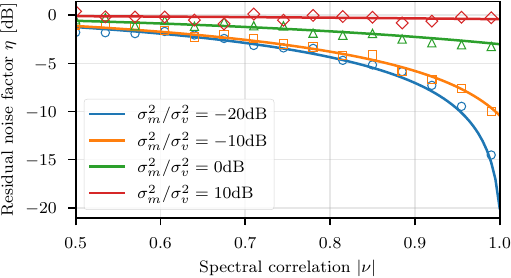}
    \caption{
    {
    Residual noise factor $\eta$ [\si{\decibel}] versus spectral correlation $|\noiseCorrTh|$ for different ratios $\interfPow/\noisePow$. 
    Solid lines:~\cref{eq:res_noise_cmvdr:factor}. 
    Markers: simulation measurements. 
    For $|\noiseCorrTh|<0.5$, all curves converge toward $\SI{0}{\decibel}$.
    }
    }
    \label{fig:theory_eta_vs_rho}
\end{figure}

\noindent
To isolate the effect of spectral correlation on the noise reduction performance, we restrict our analysis to the single-microphone case ($M=1$).
We follow a similar procedure as in~\cite{benesty_study_2010}.
In this setting, the noisy signal is given by:
\begin{align}
\scalemath{1.}{
    \v{x} = 
\begin{bmatrix*}
    \fourier{s}(\omega_k) \\
    \fourier{s}(\omega_k - \modf_{1_k})
\end{bmatrix*}
+
\begin{bmatrix*}
    \fourier{v}(\omega_k) \\
    \fourier{v}(\omega_k - \modf_{1_k})
\end{bmatrix*}.
}
\end{align}
Assuming that the target and noise are uncorrelated and that $\fourier{s}(\omega_k)$ and $\fourier{s}(\omega_k - \modf_{1_k})$ are uncorrelated,  
the noisy spectral covariance matrix is given by:
\begin{align}
\scalemath{1.}{
    \v{S}_{\v{x}} =
    \underbracket{
\begin{bmatrix*}
    \targetPow & 0\\
    0 & 0
\end{bmatrix*}
}_{\v{S}_s}
+
\underbracket{
\begin{bmatrix*}
    0 & 0\\
    0 & \interfPow
\end{bmatrix*}
}_{\interfCov}
+
\underbracket{
\noisePow
\begin{bmatrix*}
    1 & \noiseCorrTh \\
    \noiseCorrTh & 1
\end{bmatrix*}
}_{\v{S}_{\v{v}}}
}
\end{align}
where $\targetPow = \E{|\fourier{s}(\omega_k)|^2}$ is the PSD of the target, $\interfPow = \E{|\fourier{s}(\omega_k - \modf_{1_k})|^2}$ is the PSD of the modulated target component, which arises as a processing artefact and should \emph{not} be present at the output of the beamformer, $\noisePow$ is the noise PSD, and $-1 \leq \noiseCorrTh \leq 1$ denotes the spectral correlation of the noise, where $\noiseCorrTh=0$ denotes spectrally uncorrelated, therefore stationary noise, and $|\noiseCorrTh|=1$ denotes perfectly correlated (harmonic) noise.
Treating the modulated target components as interference, the overall noise covariance is $\v{S}_{\v{n}} = \interfCov + \v{S}_{\v{v}}$ and should be as small as possible after filtering.
The cMPDR minimizes $\v{w}^H \v{S}_{\v{x}} \v{w}$ subject to preserving the non-modulated target component $\fourier{s}(\omega_k)$. 
In this example, the distortionless response constraint reduces to 
\begin{align}
    \v{a}_0^H \v{w}_{\text{c}} = [1\quad 0]\,\v{w}_{\text{c}} =
    1 \iff \v{w}_{\text{c}} = [1\quad y]^T,
\end{align}
as $\v{a}_0$ is normalized being an RTF.
The optimal cMPDR beamformer is then in the form $\v{w}_{\text{c}} = [1~y]^T$.
By evaluating the output power $J = \v{w}_{\text{c}}^H \v{S}_{\v{x}} \v{w}_{\text{c}}$ and computing the Wirtinger derivative of $J$ w.r.t.~$y^*$~\cite{brandwood_complex_1983}, we find the constrained weights that give the least output power as 
\begin{align}
    \v{w}_{\text{c}} = \begin{bmatrix*}1 & -\noiseCorrTh\,\noisePow/(\noisePow + \interfPow)\end{bmatrix*}^T\hspace{-0.8em}.
\end{align}
It follows that the residual output noise power is given by
\begin{align}\label{eq:res_noise_cmvdr}
    \v{w}_{\text{c}}^H \v{S}_{\v{n}} \v{w}_{\text{c}} &= \v{w}_{\text{c}}^H (\interfCov + \v{S}_{\v{v}})  \v{w}_{\text{c}} =
    \noisePow - \noiseCorrTh^2 
    \frac{\sigma_v^4}
    {\noisePow + \interfPow}. 
\end{align}
By dividing~\cref{eq:res_noise_cmvdr} by the input noise power $\noisePow$ we obtain the relative residual noise factor $\ResNoise$ (lower is better):
\begin{align}\label{eq:res_noise_cmvdr:factor}
    \ResNoise = \frac{\v{w}_{\text{c}}^H \v{S}_{\v{n}} \v{w}_{\text{c}}}{\noisePow} = 1 - \noiseCorrTh^2 \frac{1}{1 + \interfPow / \noisePow}.
\end{align}
{
The conventional MPDR cannot operate in single-channel settings and therefore acts as a bypass, yielding $\ResNoise=1$.
With $\text{SNR}_{\text{in}} = \targetPow / \noisePow$ and the distortionless constraint preserving the target power $\targetPow$, the output SNR becomes
\begin{align}
    \text{SNR}_{\text{out}}^{\text{cMPDR}}
    =
    \frac{\text{SNR}_{\text{in}}}{\ResNoise},
    \qquad
    \text{SNR}_{\text{out}}^{\text{MPDR}}
    =
    \text{SNR}_{\text{in}}.
\end{align}
Thus, for $\noiseCorrTh \neq 0$, $\ResNoise < 1$ and the cMPDR strictly increases the output SNR.
}
It is also worth noting that, when $\interfPow = 0$,~\cref{eq:res_noise_cmvdr} matches the minimum achievable mean-squared error in estimating one complex RV from another when the two have correlation $\noiseCorrTh$ (see \cite[Eq.~33]{kasasbeh_noise_2017,adali_optimization_2014}).
{
The theoretical prediction in~\cref{eq:res_noise_cmvdr:factor} is validated in~\cref{fig:theory_eta_vs_rho}, where the statistics used in the cMPDR are estimated from 200 realizations.
}

\Cref{eq:res_noise_cmvdr:factor,fig:theory_eta_vs_rho} reveal the following properties.
First, the relative residual noise factor of the cMPDR beamformer is strictly less than 1 when $\noiseCorrTh \neq 0$.
As a result, the cMPDR improves the subband SNR, unlike the single- or multichannel Wiener filter, which does not affect it.
Second, the amount of noise reduction increases with increasing correlation. 
Noise reduction also increases as the power of the modulated target component $\interfPow$ decreases. 
In the limiting case of purely harmonic noise ($|\noiseCorrTh| = 1$) and negligible modulated target interfering power ($\interfPow \rightarrow 0$), the residual noise is exactly zero.
Third, the cMPDR becomes less effective when the modulated target component has high power, which typically occurs at higher SNRs. 
For instance, if $\interfPow = \targetPow = 100 \cdot \noisePow$ (\SI{20}{\decibel} SNR), then $\ResNoise \approx 1 - \noiseCorrTh^2 \cdot 0.01 \approx 1$, meaning the residual noise is approximately the same as for the conventional MPDR.

\subsection{Noise reduction performance: arbitrary \texorpdfstring{$\cycSize$}{Ck} and \texorpdfstring{$M$}{M}}
\label{ssec:stat:noise_red_arbitrary_ck}
\noindent
In the general case of arbitrary $\cycSize$ and $M$, the following result holds.
\begin{theorem}[Monotonicity of output power]\label{th:monotonic_ck}
Consider the cMPDR beamformer for arbitrary $\cycSize$ and $M$ at frequency $\omega_k$, with positive definite covariance matrix $\v{S}_{\v{x}}$.
Let $J^\star(\cycSize)$ denote the minimum value of the optimization problem in~\cref{eq:cmpdr_problem} for a given $\cycSize$.
Then $J^\star(\cycSize)$ is non-increasing with $\cycSize$.
\end{theorem}
\begin{proof}
The cMPDR design in~\cref{eq:cmpdr_problem} is a strictly convex quadratic program with linear equality constraint, since $\v{S}_{\v{x}}$ is positive definite.
Increasing $\cycSize$ extends the observation vector $\v{x}$ with additional frequency-shifted components while preserving the same distortionless constraint.
Hence, the feasible set for $\cycSize+1$ strictly contains the feasible set for $\cycSize$.
Since the objective $\v{w}^H \v{S}_{\v{x}} \v{w}$ is strictly convex, the minimum over a larger feasible set cannot exceed the minimum over a subset.
\end{proof}
Since the distortionless constraint preserves the desired signal power, monotonic decrease of $J^\star(\cycSize)$ directly implies monotonic decrease of the residual noise power as defined in~\cref{eq:res_noise_cmvdr}.
Consequently, the output SNR of the cMPDR improves or remains unchanged as $\cycSize$ increases.
Experimental results confirm strict improvement whenever additional correlated components are included ($\noiseCorrTh \neq 0$ and $\noisePow > 0$ in~\cref{fig:res-synth-num_mods}), in agreement with the literature~\cite{gardner_cyclic_1993}.

\section{Construction of the modulation sets \texorpdfstring{$\ModSet_k$}{}}\label{sec:calc_mod_set}\noindent
\begin{figure}[tb]
\centering
\resizebox{.95\columnwidth}{!}{%
\begin{tikzpicture}[
  >=Stealth,
  font=\small,
  line width=.5pt,
  node distance = 8mm and 8mm,  
  box/.style    = {draw, rounded corners=4pt, minimum height=7mm,
                   minimum width=23mm, align=center},
  dashedbox/.style={box, densely dashed, inner sep=1pt, inner ysep=1mm},
  dottedbox/.style={box, dotted, inner sep=1pt, inner ysep=1mm},
  input/.style   = {box, densely dashed, draw=red!70!black, fill=red!20},
  output/.style  = {box, dotted, draw=green!50!black, fill=green!20},
]

\node[coordinate] (leftanchor) {};

\node[input, right=0mm of leftanchor, anchor=west]   (noisy)  {Noisy $\v{{\bar x}}_{[N]}(n)$};
\node[box, right=of noisy] (freq)  {Freq.\ estim.\\(Sec.\ \ref{ssec:freq_est_period})};
\node[box, right=of freq]  (msc)   {Freq.\ shifts\\calc. (Sec.\ \ref{ssec:calc_shifts})};
\node[box, below=of freq]  (mod)   {Modulation\\(\cref{eq:modulation})};
\node[box, below=of mod]   (stft)  {STFT\\(\cref{eq:stft})};
\node[box, left=of stft]   (coh)   {Coherence\\filter (Sec.\ \ref{ssec:pipeline:coh_filt})};
\node[box, below=of coh]   (cov)   {Estimate\\cov.\ \& RTF};
\node[box, below=of stft]   (cmpdr) {cMPDR\\(\cref{eq:cmpdr_solution})};
\node[output, right=of cmpdr] (clean) {Clean $\hat{s}(\omega_k)$};

\newcommand{\smallvdots}{\vbox{\baselineskip=3pt \lineskiplimit=0pt
\hbox{.}\hbox{.}\hbox{.}}}
\node[dottedbox, below=of noisy, yshift=3mm] (leftset)
    {$\smallvdots$\\$\bar{\v{x}}_{[N]}(n)e^{jn\modf}$\\[0.3em]$\smallvdots$};

\node[dottedbox,
    right=of stft] (rightset)
    {$\smallvdots$\\$\v{x}(\omega_k-\modf)$\\[0.3em]$\smallvdots$};

\draw[->] (noisy) -- (freq);
\draw[->] (noisy) to (mod);

\draw[->] (freq) -- node[above]{$\cSetEst$} (msc);

\draw[->] (msc) to node[below, fill=white, inner sep=1pt, yshift=4pt]{Cand.\ set $\CandSet^{\Delta}$} (mod);

\draw[->] (mod) -- (stft);

\draw[dotted, -] (mod.south) to [out=-90, in=-60] (leftset.east);
\draw[dotted, -] (stft.south) to [out=-90, in=+210] (rightset.west);

\draw[->] (stft) -- (coh);
\draw[->] (stft) -- (cov);
\draw[->] (stft) -- (cmpdr);
\draw[->] (coh) -- node[left]{$\ModSet_k$} (cov);

\draw[->] (cov) -- (cmpdr);
\draw[->] (cmpdr) -- (clean);
\end{tikzpicture}
}
%
\vspace{1pt}
\caption{Overview of the proposed method: harmonic frequency estimation, signal modulation, STFT, coherence filtering, and cMPDR beamforming.}
\label{fig:cmpdr_scheme}
\end{figure}%
The multi-band signal $\v{x}(\ModSet_k, \omega_k)$ serving as input to the cMPDR beamformer at frequency bin $k$ is constructed by modulating the noisy input with the $\cycSize$ frequencies in the modulation set $\ModSet_k$, defined in~\cref{eq:mod_set_definition}.
As discussed in~\cref{ssec:stat:noise_red_corr}, the output SNR per subband is maximized when the modulation frequencies induce the strongest spectral correlation.
In principle, this would require testing all possible frequencies and selecting those that maximize the correlation; however, such an exhaustive search is computationally infeasible.
Instead, candidate frequency shifts are selected following established principles from the literature, which suggest including all combinations of cyclic frequencies of target and noise signals~\cite{gardner_cyclic_1993,zhang_reduced-rank_2006}.
To this end, we identify candidate shifts across all estimated resonant frequencies, accounting for potential estimation errors or missed components in the frequency-finding stage (\cref{ssec:freq_est_period}), and collect them in a candidate set. 
For each frequency bin $k$, the candidate modulations whose spectral coherence exceeds a threshold are included in the modulation set $\ModSet_k$, keeping at most the $\cmax$ most coherent candidates.

Specifically, resonant noise frequencies are first estimated (\cref{ssec:freq_est_period}) and used to form the candidate set (\cref{ssec:calc_shifts}). 
The noisy signal is then modulated and transformed to the STFT domain, after which coherence filtering (\cref{ssec:pipeline:coh_filt}) identifies the most informative shifts.
Based on these selected modulations, spectral-spatial covariance matrices and RTFs are estimated, and the beamforming weights are computed and applied to recover the denoised signal.
The overall procedure is summarized in~\cref{fig:cmpdr_scheme}.

\subsection{Estimation of resonant frequencies}\label{ssec:freq_est_period}\noindent
The spectral peak frequencies of the noise, referred to as resonant frequencies, are estimated using the unmodified periodogram method \cite[Ch.~4]{stoica_spectral_2005}.
Candidate cyclic frequencies are later obtained from their pairwise differences.
A rectangular window is preferred over smooth windows because, although smooth windows reduce variance, they broaden the main lobe and lower spectral resolution.
Since CS-based methods are highly sensitive to frequency estimation errors~\cite{ojeda_adaptive-fresh_2010,bologni_cyclic_2025}, maximizing spectral resolution is critical, motivating the choice of a single, long rectangular window over a Bartlett estimate with shorter windows and reduced variance. 
The periodogram is computed as the magnitude-squared DFT of the noise-only signal at the reference microphone over $N_v$ samples:
\begin{align}
    |\operatorname{DFT}(\bar{v}_0(n))|^2,\qquad n=0,\ldots,N_v -1.
\end{align}
To improve frequency resolution, the DFT uses $K_v > N_v$ points with zero-padding, yielding an interpolated periodogram.
In practice, $K_v$ should be chosen much larger than the STFT frame length $K$ to ensure accurate estimation of the harmonic frequencies.
Spectral peaks are located with \texttt{find\_peaks} (SciPy 1.15.1), subject to constraints on minimum and maximum frequency, minimum peak distance, peak height ratio, and a maximum count.
The top $\csizeest$ peaks in amplitude are selected as resonant frequency estimates and collected in a set $\cSetEst = \{\hat{\alpha}_{\cidxest}\}_{\cidxest=1}^{\csizeest}$.
Assuming the noise resonant frequencies remain constant, the periodogram estimation is performed only once per audio recording.
Although resonant frequency estimates are most accurate when obtained from noise-only recordings, satisfactory performance can still be achieved using the noisy mixture, especially in low-SNR scenarios.

\subsection{Calculation of frequency shifts}\label{ssec:calc_shifts}\noindent
Based on the estimated locations of the noise harmonics, we construct a set of candidate frequency shifts $\CandSet$ to be applied to the signal.
Positive shifts move the signal content upwards in frequency.
As alluded to in~\cref{fig:harmonic_components}, the goal is to obtain frequency-shifted versions of the signal that are strongly correlated with the original, thereby enabling spectral beamforming.
We consider two alternative constructions of $\CandSet$: $\CandSet^{\times}$, the integer-multiple set, which assumes a perfectly harmonic or CS model, and $\CandSet^{\Delta}$, the difference-based set, which accommodates inharmonic or ACS processes.
\subsubsection{Integer-multiple shifts (\texorpdfstring{$\times$}{x} strategy)}\label{sssec:down_shift}\noindent
In our previous work~\cite{bologni_cyclic_2025}, the frequency shifts were computed based on the assumption that all harmonic components lie exactly at integer multiples of the estimated fundamental frequency $\hat{\alpha}_1$, yielding the following candidate modulation set:
\begin{align}\label{eq:mod_set_down}
     \CandSet^{\times} = \{ -r\hat{\alpha}_1 \}_{r=0}^{\csizeest-1}.
\end{align}
While simple, this approach has two key limitations. 
First, real-world acoustic signals often deviate from the ideal harmonic model, as discussed in~\cref{sec:intro}, and such model mismatch degrades the performance of CS-based methods. 
Second, the method only considers negative frequency shifts, whereas including both negative and positive modulations is generally beneficial for performance.
\begin{figure}[bt]
    \centering
    \includegraphics[width=0.85\linewidth]{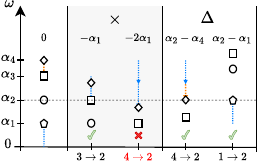}
    \caption{
    Comparison of frequency-shift calculation strategies for aligning harmonics to working frequency $\alpha_2$.
    Column 1: original spectrum.
    Columns 2-3: integer-multiple downward shifts (\cref{sssec:down_shift});  
    Columns 4-5: difference-based shifts (\cref{sssec:diff_shift}).  
    Markers distinguish individual harmonics.
    }
    \label{fig:mod_sets}
\end{figure}%

\subsubsection{Difference-based shifts (\texorpdfstring{$\Delta$}{Delta} strategy)}\label{sssec:diff_shift}\noindent
To address the limitations of the integer-multiple approach, we propose a difference-based construction of the candidate set. 
The modulations are computed as all pairwise differences between the estimated resonant frequencies in $\cSetEst$, accounting for deviations from the ideal harmonic model:
\begin{align}\label{eq:mod_set_diff_delta}
    \CandSet^{\Delta} = \{ \hat{\alpha}_\cidxest - \hat{\alpha}_{r} \}_{\cidxest,r=1}^{\csizeest}
\end{align}
This strategy produces both positive and negative frequency shifts and remains effective even when the harmonics are arbitrarily far from the integer multiples of the fundamental frequency.

\Cref{fig:mod_sets} compares the two modulation strategies for a working frequency $\alpha_2$.
The first column shows the original spectrum; in this example, $\alpha_2$ and $\alpha_3$ lie at exact integer multiples of the fundamental $\alpha_1$, while $\alpha_4$ is placed off the harmonic grid to illustrate inharmonicity.
Columns 2-3 apply the downward shifts from \eqref{eq:mod_set_down}, which only align exact harmonics.
As seen with $\alpha_4$ in column 3 in this example, any deviation from the ideal model causes the alignment to fail.
In contrast, columns 4-5 use the difference-based shifts from \eqref{eq:mod_set_diff_delta}, successfully aligning $\alpha_4$ and $\alpha_1$ to $\alpha_2$.
This confirms that the difference-based approach yields correct alignment for arbitrary resonant frequencies.

\Cref{fig:synth-inharmonicity} shows SI-SDR results for the benchmark MPDR and the proposed cMPDR as a function of the noise inharmonicity.
Higher inharmonicity means that the resonant frequencies deviate more from integer multiples of the fundamental.
The cMPDR uses either the $\times$ or $\Delta$ strategy to compute frequency shifts, assuming the resonant frequencies are known in this toy example.
Additional details are given in~\cref{sec:exp_setup}.
When the noise is perfectly harmonic, both strategies perform equally well, and cMPDR achieves a \SI{10}{\decibel} gain over the MPDR.
As the inharmonicity increases, the $\times$ strategy---based on integer-multiple assumptions---rapidly degrades, with no gains for as little as $0.5\%$ inharmonicity, highlighting the sensitivity of strict harmonic models to deviations from perfect periodicity.
In contrast, the $\Delta$ strategy remains effective across all inharmonicity levels by aligning shifts with the observed spacing between resonant frequencies.
The $\Delta$ strategy is therefore adopted in all the experiments that follow.

\subsection{Coherence-based shift selection}\label{ssec:pipeline:coh_filt}\noindent
\Cref{ssec:calc_shifts} described two methods for identifying candidate modulations based on the estimated cyclic frequencies.
This section introduces \Cref{alg:coherence_filtering} to select the optimal candidates for inclusion in the final modulation sets based on spectral coherence.
Spectral coherence, defined in~\cref{eq:coherence_single_ch}, quantifies the correlation between original and modulated signals at each frequency bin, and has been used to detect CS signals in noise~\cite{kim_cyclostationary_2007,yang_cyclostationary_2019}.
As demonstrated in~\cref{ssec:stat:noise_red_corr}, highly correlated components lead to improved noise suppression, making coherence an effective selection criterion.

The selection process applies two sequential filtering stages to reduce computational load.
First, modulations must exceed a coherence threshold to qualify for inclusion.
Second, when the number of qualifying modulations exceeds the user-specified maximum, the algorithm retains those with the highest coherence values.
By convention, $\ModSet_k = \{0\}$ for $k \notin [K_{\min}, K_{\max})$, the bin indices corresponding to the admissible frequency range, so that cMPDR reduces exactly to MPDR outside this range.
The parameter values used in the implementation, including the admissible frequency range, are listed in~\cref{tab:alg_params}.
\begin{algorithm}[bt]
\DontPrintSemicolon
\SetNoFillComment
\LinesNotNumbered 
\SetKwFor{For}{for}{}{}
\SetKwIF{If}{ElseIf}{Else}{if}{}{elif}{else}{}
\caption{Coherence-based shift selection}
\label{alg:coherence_filtering}

\KwIn{Candidate set $\CandSet^{\Delta}$,
 noisy $\fourier{x}(\omega_k)$, coherence threshold $\cohThreshold$, max modulations $\cmax$}
\KwOut{Modulation sets $\ModSet_k,~k \in [K_{\min},K_{\max})$}
\ForEach{$\modf \in \CandSet^{\Delta}$}{
  \For{$k = K_{\min}$ \KwTo $K_{\max}-1$}{
    Compute coherence $\hat{\gamma}_x(\modf,\omega_{k})$ (\cref{eq:acp_estimator:coh})
  }
}
\For{$k = K_{\min}$ \KwTo $K_{\max}-1$}{
  $\ModSet_k \gets \{\modf :  \hat{\gamma}_x(\modf,\omega_{k}) \geq \cohThreshold\} \cup \{0\}$\;
  $\ModSet_k \gets$ keep $\cmax$ most coherent items in $\ModSet_k$ 
}
\end{algorithm}

\section{Experimental setup}\label{sec:exp_setup}\noindent
\subsection{Parameter estimation and algorithms}\noindent
Computing the beamforming weights requires estimates of the noisy spectral-spatial covariance matrix, as well as the RTFs, all of which are unknown in practice.
The matrix $\hat{\v{S}}_{\v{x}}$ is estimated from the noisy measurements using the ACP.
Recursive averaging 
is applied to update $\hat{\v{S}}_{\v{x}}$ at each STFT frame.
Diagonal loading is adopted to improve the conditioning of $\hat{\v{S}}_{\v{x}}$ and reduce the sensitivity of the beamformer to errors in the estimated RTF vector $\v{a}$~\cite{cox_robust_1987}.
Specifically, a scaled identity matrix is added to the noisy covariance matrix, with the scaling chosen to bound the condition number of the loaded matrix~\cite{moore_compact_2022}.
The RTFs $\v{a}(\omega_k)$ are estimated using the covariance whitening (CW) technique~\cite{markovich-golan_performance_2015}.
The noise covariance matrix $\hat{\v{S}}_{\v{v}}$, required to perform CW, is estimated from a separate noise-only segment.
The conventional MPDR beamformer is used as a benchmark; its spatial covariance matrices are estimated via standard recursive averaging with diagonal loading, and its RTFs are obtained using the same CW procedure as for the proposed method.
To assess the impact of RTF estimation on performance, we also evaluate oracle versions of the beamformers, denoted as MPDR+ and cMPDR+, that assume access to the true RTFs.

Frequency estimation (\cref{ssec:freq_est_period}) is performed on the same noise-only realization used to estimate the covariance matrices.
For peak selection after frequency estimation with the periodogram, we enforce a minimum separation between peaks and limit both the peak amplitude ratio and the maximum number of detected peaks.
The admissible frequency range is set to \SI{20}{\hertz} to \SI{2.5}{\kilo\hertz}, as preliminary experiments showed that higher resonant frequencies are difficult to estimate with sufficient accuracy.
Coherence filtering (\cref{ssec:pipeline:coh_filt}) is based on the noisy data.
Estimation and algorithmic parameters are reported in~\cref{tab:alg_params}.
\begin{table}[bt]
\centering
\small
\caption{{Estimation and algorithmic parameters.}}
\label{tab:alg_params}
\begin{tabular}{ll}
\toprule
Parameter & Value \\
\midrule
DFT size (STFT) & $K=2048$ \\
STFT window & Square-root Hann \\
STFT overlap & $75\%$ \\
Covariance smoothing constant & $\beta_x=0.95$ \\
Max condition number & $\kappa_0 = 1000$ \\
\midrule
Minimum peak distance & \SI{20}{\hertz} \\
Maximum peak ratio & $\num{e4}$ \\
Maximum number of peaks & $\csizeest = 20$ \\
Periodogram DFT size & $K_v = 2^{17}$ \\
Admissible frequency range & \SI{20}{\hertz} -- \SI{2.5}{\kilo\hertz} \\
\midrule
PSD dynamic range & $D_{\text{PSD}}=1000$ \\
Coherence threshold & $\cohThreshold=0.6$ \\
Maximum retained shifts & $\cmax=8$ \\
\bottomrule
\end{tabular}
\end{table}
\subsection{Simulation parameters and metrics}\label{ssec:impl:settings}\noindent
Performance is evaluated using SI-SDR for both simulated and real data~\cite{roux_sdr_2019}.
For real recordings, we additionally report short-time objective intelligibility (STOI) scores~\cite{taal_algorithm_2011}.
PESQ was not included, as at low SNRs scores for all methods saturate to a floor level~\cite{rix_perceptual_2001}.
All results are shown as score improvements, computed by subtracting the score of the unprocessed signal at the first microphone from the score of each beamformer.

We simulate a target speech source and an interfering noise source, both modeled as omni-directional point sources.
Spatially uncorrelated white Gaussian noise is added to simulate microphone self-noise.
Speech signals are recordings of the Harvard sentences, uttered by either a male or a female speaker~\cite{noauthor_ieee_1969}.
The room impulse responses (RIRs) for the target and interferer are randomly selected from the Bar-Ilan dataset~\cite{hadad_multichannel_2014}.
For synthetic noise experiments, we select from 26 RIRs with $\text{RT60} = \SI{0.61}{\second}$, while for real noise experiments we use 26 RIRs with $\text{RT60} = \SI{0.16}{\second}$, unless otherwise stated.
The angles and the distances between the point sources and the array vary independently at each Monte Carlo trial.
Results are averaged over multiple Monte Carlo runs, each using different RIRs, noise realizations, and target signals.
For each run, a speech segment and a noise segment are randomly selected and weighted to obtain the desired SNR, which is defined with respect to the target and noise powers at the reference microphone, and then summed to form a mixture of the desired length.
Lines in the plots indicate mean values, and shaded areas correspond to 95\% confidence intervals.
The default simulation parameters are summarized in~\cref{tab:sim_params}.
\begin{table}[bt]
\centering
\small
\caption{{Simulation and data generation parameters.}}
\label{tab:sim_params}
\begin{tabular}{ll}
\toprule
Parameter & Value \\
\midrule
Sampling frequency & $f_s=\SI{16}{\kilo\hertz}$ \\
Number of microphones & $M=2$ \\
Inter-microphone spacing & \SI{8}{\centi\meter} \\
Noisy sequence duration & \SI{2}{\second} \\
Noise-only sequence duration & \SI{2}{\second} \\
Interferer SNR & \SI{-10}{\decibel} \\
Self-noise SNR & \SI{30}{\decibel} \\
Noise spectral correlation & $\noiseCorr=0.8$\\
Monte Carlo runs & $50$ \\
\bottomrule
\end{tabular}
\end{table}

\subsection{Noise datasets}\label{ssec:datasets}\noindent
The cMPDR is evaluated against three different noise types: synthetic, CS noise; real, ACS motor noise; real, mostly non-stationary environmental noise.
\subsubsection{Synthetic harmonic noise}\label{sss:synth}\noindent
The synthetic CS noise is modeled as a random harmonic signal with adjustable correlation across harmonic components.
The model allows controlled variation of the correlation level, reflecting the partial correlation typically observed in real noise. 
Formally, let
$\proc{{\bar u}(n), n \in \mathbb{Z}}$ be a CS process where a scalar parameter $\noiseCorr$ controls whether all harmonics share the same amplitude variation over time or vary independently:
\begin{subequations}
\begin{align}
{\bar u}(n) &= \sum_{p=1}^{P_u} {\bar a}_p(n; \noiseCorr) \cos{(\omega_p n  + \phi_p)},\label{eq:harm_noise_model}\\
{\bar a}_p(n; \noiseCorr) &= \left[\noiseCorr {\bar b}(n) + (1-\noiseCorr){\bar c}_p(n)\right] d_p,
\end{align}
\end{subequations}
where $\omega_p = \omega_0 p$.
The frequency $\omega_0$ is a RV drawn from $2\pi \cdot \mathcal{U}(60, 150)$, and the phases $\phi_p$ are independently drawn from $\mathcal{U}(-\pi, \pi)$.
The processes $\proc{{\bar b}(n)}$ and $\proc{{\bar c}_p(n)}$ are WSS and describe the temporal amplitude fluctuations. 
The amplitudes $d_p$ are independent RVs drawn from $\mathcal{U}(1, 10)$. 
Both $\proc{{\bar b}(n)}$ and $\proc{{\bar c}_p(n)}$ consist of independent Gaussian RVs distributed as $\mathcal{N}(0, 10)$ and lowpass filtered by a 4th order Butterworth filter with cutoff frequency $f_c = \SI{5}{\hertz}$.
The parameter $\noiseCorr \in [0, 1]$ controls the correlation of the amplitude envelopes across components, where $\noiseCorr$ closer to 1 gives more spectral correlation.
The spectral correlation is set to $\noiseCorr=0.8$ unless specified differently.
The number of components is set to $P_u = 16$.

{
\subsubsection{Real UAV motor noise (DREGON dataset)}\label{sss:dregon}\noindent
The second noise set is a subset of the DREGON dataset~\cite{strauss_dregon_2018} that consists of 21 motor-only recordings from a quadrotor unmanned aerial vehicle (UAV, commonly known as drone) equipped with an 8-channel cube-shaped microphone array.
The UAV was suspended on a nylon string and held stationary to capture pure motor noise under low-reverberation conditions.
The recordings include individual propeller noise at different rotational speeds, as well as combined engine noise.
The initial and final \SI{10}{\second} of each recording are trimmed to remove quiet segments and transients from motor start-up and shutdown.
This noise set exhibits pseudo-harmonic, ACS noise, with spectral peaks at approximate multiples of the motor rotation frequency, making it the primary real-world benchmark for cMPDR.

\subsubsection{Real non-stationary noise (MUSAN dataset)}\label{sss:musan}\noindent
The third noise set consists of recordings drawn from the Freesound portion of the MUSAN corpus~\cite{snyder_musan_2015}, as adapted by Ko \emph{et al.\ }\cite{ko_study_2017}.
We used the public OpenSLR release of this dataset~\cite{openslr28}.
It comprises a wide range of 843 mostly non-stationary noise signals that are not aligned with the assumptions underlying cMPDR.
This dataset is used to evaluate behavior outside the targeted ACS setting.
In these conditions, the cyclic advantage is expected to vanish, such that the cMPDR effectively reduces to MPDR and does not incur additional performance loss.
}

\section{Experimental results}\label{sec:results}\noindent
Audio examples corresponding to the results in this section are available online.\footnote{\url{https://screeen.github.io/cmpdr-demo/}}
\Cref{fig:spec_examples_good} provides a qualitative example of the noise suppression achieved by the proposed method.
In the presence of strong tonal components, the cMPDR can better attenuate structured low-frequency interference than the MPDR, resulting in clearer speech.
The following subsections provide systematic quantitative evaluations across three noise datasets.
\begin{figure}[tbp]
    \input{pics-musan/spectrograms}
    \caption{
    Mel spectrograms of clean speech mixed with buzz noise from the MUSAN recording \texttt{noise-free-sound-0002} (\SI{-5}{\decibel} SNR), showing the noisy input, the clean target signal, and the outputs of the MPDR and proposed cMPDR. 
    The highlights show that the cMPDR can suppress strong tonal noise thanks to cyclostationarity-aware processing.
    }
    \label{fig:spec_examples_good}
\end{figure}%

\subsection{Synthetic harmonic noise experiments }\label{ssec:experiments_synth}\noindent
This section compares the performance of the MPDR and cMPDR beamformers on the synthetic noise set, which features random harmonic signals with adjustable spectral correlation (\cref{sss:synth}).
\begin{figure}[tbp]
    \input{results_synthetic_inharm}
    \caption{
    Comparison of the $\times$ and the $\Delta$ strategies for calculating frequency shifts using synthetic harmonic noise.
    Default parameters are given in~\cref{sss:synth} and~\cref{tab:sim_params}; here, resonant frequencies are assumed known and only inharmonicity is varied.
    }
    \label{fig:synth-inharmonicity}
\end{figure}
\begin{figure}[tbp]
\input{results_synthetic}%
\caption{Synthetic harmonic noise experiments. 
SI-SDR improvements as a function of (a) spectral correlation, (b) maximum number of modulations, (c) number of microphones, (d) interferer SNR, (e) reverberation time, (f) frequency estimation error, and (g) distance between the fundamental frequencies of target and noise signals.
Unless otherwise specified, parameters are set to the default values in~\cref{tab:sim_params}.
Oracle variants (``+") use the true RTF.
}
\label{fig:res_synth}
\end{figure}%

\Cref{fig:synth-inharmonicity}, already discussed in~\cref{ssec:calc_shifts}, compares the $\times$ and the $\Delta$ strategies for computing frequency shifts under varying noise inharmonicity, assuming known resonant frequencies.
These are defined as $\dot{\omega}_p = \omega_0 p (1 + \omega_{\text{err}} / 100)$ for $p>1$, while the fundamental is kept fixed at $\dot{\omega}_1=\omega_0$.
The performance of the $\Delta$ strategy remains stable across all tested inharmonicity levels and is therefore used in all subsequent experiments.

\Cref{fig:res_synth} reports the output SI-SDR on speech corrupted by synthetic noise.
In~\crefrange{fig:res-synth-beta}{fig:res-synth-rt60}, resonant frequencies are estimated via the periodogram.
\Cref{fig:res-synth-beta} shows that the performance of the proposed cMPDR improves with increasing noise correlation $\noiseCorr$.
For $\noiseCorr > 0.8$, it even outperforms the oracle MPDR, in agreement with the theoretical result in~\cref{ssec:stat:noise_red_corr}, which favors cMPDR for highly correlated noise.
\Cref{fig:res-synth-num_mods} shows the effect of $C_{\text{max}}$, the maximum number of modulations per set.
With $C_{\text{max}} = 2$, the cMPDR already achieves a \SI{5}{\decibel} gain over the MPDR.
Increasing $C_{\text{max}}$ further improves SI-SDR,
providing empirical support for~\cref{th:monotonic_ck}.
\Cref{fig:res-synth-m} shows that, unlike the other methods, cMPDR performance remains roughly constant as the number of microphones increases: because the noise is highly correlated ($\noiseCorr=0.8$), spectral cues already dominate over spatial cues, and cMPDR performs well even with a single microphone.
This behavior is specific to the high-correlation regime considered here: cMPDR does benefit from additional microphones in the real-data experiments (\cref{fig:dregon-160-sisdr-m,fig:dregon-160-stoi-m}) and in additional synthetic experiments with $\noiseCorr=0.65$ (not shown).
\Cref{fig:res-synth-snr} presents SI-SDR as a function of interferer SNR (iSNR).
In the low-SNR regime, both the proposed and oracle cMPDR outperform the oracle MPDR, confirming their robustness.
At iSNRs above \SI{10}{\decibel}, the cyclic methods perform similarly to the benchmark, consistent with~\cref{eq:res_noise_cmvdr:factor}.
Next,~\cref{fig:res-synth-rt60} shows that increasing the reverberation time degrades the performance of all methods.
However, the proposed cMPDR is less affected than the MPDR: its SI-SDR decreases by about \SI{2.5}{\decibel}, compared with approximately \SI{4.5}{\decibel} for the MPDR, suggesting that the proposed method is more robust to reverberation.

\Cref{fig:res-synth-mod-err} evaluates the sensitivity of the proposed beamformer to errors in the estimated resonant frequencies.
Unlike~\cref{fig:synth-inharmonicity}, which assumes access to true but increasingly inharmonic frequencies, here the true resonant frequencies are perturbed before computing the candidate shifts using
$
\ddot{\omega}_p = \omega_0 p (1 + \omega_{\text{err}} / 100 \cdot \mathcal{U}(-1, 1))
$.
MPDR and MPDR+ do not use frequency information and are therefore not affected.
On the other hand, cMPDR outperforms MPDR in SI-SDR by \SI{8}{\decibel} with perfect estimation and by \SI{4}{\decibel} with 1\% error, while converging to MPDR performance at 10\% error.

Lastly,~\cref{fig:res-synth-fun-freq-spectral-distance} investigates the effect of increasing overlap between the resonant frequencies of the target and interferer.
For this experiment only, the target is also modeled as a synthetic harmonic signal following~\cref{eq:harm_noise_model}, rather than speech, with a fixed fundamental frequency of \SI{100}{\hertz}.
The interferer fundamental frequency is then swept between \SIrange{95}{105}{\hertz}, progressively increasing the overlap between the harmonic structures of the target and interferer.
The true resonant frequencies are used for the calculation of frequency shifts to isolate the effect of this overlap.
Since frequency modulations are applied in the time domain (\cref{eq:modulation}), the cMPDR can resolve frequency separations smaller than the STFT frequency resolution of \SI{7.8}{\hertz}.
Accordingly, for fundamental frequency distances as small as \SI{5}{\hertz}, the proposed cMPDR outperforms the MPDR by approximately \SI{8}{\decibel}.
As the difference between the resonant frequencies of the target and interferer decreases, performance gradually degrades, with a sharp drop occurring only below \SI{0.25}{\hertz}.

\subsection{Real UAV motor noise experiments (DREGON dataset)}\label{ssec:exp_real_dregon}\noindent
\begin{figure*}[!t]
    \input{results_real_dregon_160}
    \caption{{Real UAV motor noise experiments (DREGON dataset). Improvements in terms of SI-SDR and STOI as a function of maximum number of modulations, number of microphones, interferer SNR and RT60.
    Unless otherwise specified, parameters are set to the default values in~\cref{tab:sim_params}.
    Oracle variants (``+") use the true RTF.
    }}
    \label{fig:dregon-160}
\end{figure*}%
This section analyzes performance in recovering target speech buried in noise from UAV motor recordings.
Each row of~\cref{fig:dregon-160} corresponds to a different evaluation metric, and each column to a different parameter.
In terms of SI-SDR, increasing the number of modulating frequencies generally improves performance, with gains up to \SI{5.2}{\decibel} for $C_{\text{max}} = 16$ (\cref{fig:dregon-160-sisdr-mods}).
More microphones are also beneficial, though gains are smaller for methods relying on CW to estimate RTFs compared to the ``$+$" variants that use oracle RTFs (\cref{fig:dregon-160-sisdr-m}).
As in the synthetic experiments, the cMPDR shows clear advantages in the low-SNR regime, while matching the MPDR performance for iSNRs $\geq 0$ (\cref{fig:dregon-160-sisdr-snr}).
Performance under increasing reverberation time (RT60) is shown in~\cref{fig:dregon-160-sisdr-rt60}, with all methods degrading with higher RT60.
STOI trends follow those of SI-SDR closely, with cMPDR achieving up to 0.06 improvement over the MPDR (\crefrange{fig:dregon-160-stoi-mods}{fig:dregon-160-stoi-rt60}).
These gains remain modest in absolute terms because cMPDR processing is confined to the \SI{20}{\hertz}--\SI{2.5}{\kilo\hertz} range (\cref{ssec:pipeline:coh_filt}); above this range the beamformer coincides with the MPDR, so noise components at higher frequencies receive only spatial, not spectral, suppression, limiting the achievable STOI improvement.
Interestingly, increasing $C_{\text{max}}$ beyond 4 does not lead to consistent improvements in either SI-SDR or STOI, suggesting that most of the benefit from additional cyclic components is captured at relatively low values.

{
\subsection{Real non-stationary noise experiments (MUSAN dataset)}\label{ssec:exp_real_musan}\noindent
\begin{figure}[tbp]
    \input{results_real_musan}
    \caption{{Real non-stationary noise experiments (MUSAN dataset). 
    SI-SDR improvements as a function of number of microphones and interferer SNR.
    Unless otherwise specified, parameters are set to the default values in~\cref{tab:sim_params}.
    Oracle variants (``+") use the true RTF.
    }}
    \label{fig:musan}
\end{figure}%
\Cref{fig:musan} reports SI-SDR and STOI improvements under non-stationary environmental noise from the MUSAN dataset, as a function of the number of microphones $M$ and interferer SNR.
In contrast to the ACS scenarios in~\cref{ssec:experiments_synth,ssec:exp_real_dregon}, MPDR and cMPDR exhibit nearly identical performance across all tested conditions. 
The oracle variants MPDR+ and cMPDR+ also approximately coincide for both SI-SDR and STOI.
Cyclic processing does not offer consistent benefits in this setting.
In most MUSAN recordings, the absence of persistent spectral correlation keeps the coherence of candidate shifts below $\cohThreshold$, leading to their exclusion from the modulation sets.
As a result, the spatial-spectral covariance matrix estimate reduces to the conventional spatial covariance matrix, and cMPDR reduces to MPDR.
These results confirm that the proposed method does not suffer performance losses when the noise deviates from the ACS model.
When cyclic structure is absent, the coherence-based selection mechanism effectively reverts the algorithm to standard MPDR behavior, preserving baseline performance under general non-stationary interference.}

\subsection{Runtime comparison}\label{ssec:exp_runtime}\noindent
The final experiment compares the runtime of the proposed cMPDR with the conventional MPDR.
Results, reported in~\cref{tab:runtime_comp}, are averaged over 50 runs per dataset in a single thread following one warm-up run on an Apple M1 Pro CPU using our publicly available Python implementation~\cite{github_2026}.
The measured runtime covers the full pipeline in~\cref{fig:cmpdr_scheme}: estimation of resonant frequencies, covariance matrices, and RTF vectors, modulations, coherence filtering, and beamforming.
Simulation parameters follow from~\cref{tab:alg_params,tab:sim_params}.

On the two real-data datasets, the runtime overhead is limited: 33--39\% over MPDR, in accordance with the small fraction of frequency bins classified as coherent (1.6--1.8\%, \ie 16--18 out of $K/2+1=1025$ bins).
For coherent bins, the averaged modulation set size $\cycSize$ remains well below $\cmax$, and the averaged coherence value $\hat{\gamma}_x = 0.77$ on the DREGON dataset indicates strong spectral correlation.
On the MUSAN dataset, although roughly the same number of bins as in DREGON are deemed coherent, $\hat{\gamma}_x$ is as low as 0.61, barely above the threshold $\cohThreshold$, due to impulsive sounds such as impacts or bursts that briefly produce broadband coherence.
These false positives add computational overhead but do not significantly affect output results, as shown in~\cref{fig:musan}.
The synthetic dataset, with its higher spectral correlation, exhibits a larger average set size (6.03) and a higher proportion of coherent bins (5.1\%), resulting in the highest slowdown ($1.47\times$).

\begin{table}[bt]
\centering
\small
\caption{Runtime of the proposed cMPDR relative to MPDR. 
Average $\cycSize$ and $\hat{\gamma}_x(\modf,\omega_{k})$ are computed over coherent bins only.}
\label{tab:runtime_comp}
\begin{tabular}{lcccc}
\toprule
Dataset & Slowdown ($\downarrow$) & $\cycSize$ & $\hat{\gamma}_x$ & Coh.~bins \\
\midrule
Synthetic & $1.47\times$ & 6.03 & 0.78 & 5.1\% \\
DREGON    & $1.33\times$ & 4.29 & 0.77 & 1.6\% \\
MUSAN     & $1.39\times$ & 3.22 & 0.61 & 1.8\% \\
\end{tabular}
\end{table}

{\subsection{Discussion}\label{ssec:discussion}\noindent}
\subsubsection{Computational complexity}

Computing the cMPDR weights requires inversion of a covariance matrix of dimension $M\cycSize$, resulting, in the worst case, in complexity $O((M\cycSize)^3)$.
In practice, the increased dimensionality occurs only for frequency bins exhibiting 
sufficiently high spectral coherence; for all remaining bins, $\cycSize=1$ and the cMPDR reduces to the conventional MPDR with identical complexity.
The empirical impact of this is quantified in~\cref{ssec:exp_runtime}.

\subsubsection{Behavior in non-stationary noise}
The proposed cMPDR does not fail in non-stationary noise conditions.
The coherence-based shift selection discards frequency shifts that give weak spectral coherence, and the method reduces to conventional MPDR.
Matrix conditioning is stabilized by diagonal loading (\cref{sec:exp_setup}) and no instability is observed in the evaluated scenarios.

However, the current implementation is not adaptive to varying harmonic structures, and is most effective when the interferer exhibits a stable spectral profile (\eg a motor at approximately constant speed).
Tracking time-varying harmonic patterns is fundamentally more challenging than tracking narrowband spatial covariance matrices. 
Whereas spatial covariance matrices scale approximately by a scalar factor when sources are stationary, spectral covariance matrices are tied to the instantaneous periodicities of the signal and change with every shift in operating regime (\eg motor speed or load).
Developing adaptive extensions for such scenarios remains an important direction for future work.

\subsubsection{Coherence-based selection}
A related limitation concerns the coherence-based shift selection criterion.
As noted in~\cref{ssec:exp_runtime}, impulsive sounds can produce momentary broadband spectral coherence that, when averaged over the recording, pushes the estimated $\hat{\gamma}_x$ above the threshold $\cohThreshold$ for some bins.
The proposed coherence-based shift selection method (\cref{alg:coherence_filtering}) does not distinguish correlation caused by persistent (A)CS structure from that caused by transient broadband impulses, leading to unnecessary computation without a corresponding quality benefit.
Designing a method that is robust to such transients remains an open problem.

As with conventional MPDR/MVDR distortionless beamformers, the cMPDR would typically be combined with a post-processing stage in a complete speech-enhancement system. 
Here, we consider the beamformer in isolation to highlight its behavior, but further noise suppression can be achieved by subsequent processing~\cite{bologni_two-step_2026}.

\section{Conclusion}\label{sec:conclusions}\noindent
This work proposed the cMPDR beamformer, extending the MPDR framework by exploiting both spatial and spectral correlations of almost-cyclostationary interference. 
Two mechanisms enable this: a difference-based shift calculation that derives modulation frequencies from pairwise differences between estimated resonant frequencies, avoiding assumptions of a rigid harmonic grid; and a coherence-based shift selection that removes weakly correlated spectral components, allowing the method to reduce to conventional MPDR when spectral redundancy is absent.

Theoretical analysis established that residual noise power decreases monotonically with the number of coherent cyclic components, and that the performance advantage over spatial-only beamforming increases monotonically with the degree of spectral correlation.
Experiments on synthetic harmonic noise, real UAV motor recordings, and non-stationary noise confirmed these predictions: the cMPDR achieves consistent SI-SDR and STOI improvements over the MPDR benchmark, with gains of up to \SI{5}{\decibel} in low-SNR conditions, while matching MPDR performance when the noise lacks persistent spectral structure.
The method is most effective for interferers with stable spectral properties.
Extending the framework to track time-varying harmonic structures remains an important direction for future work.
\bibliographystyle{IEEEtran}
\bibliography{IEEEabrv,references}
\end{document}